\documentclass[superscriptaddress,article,twocolumn,pra,showpacs,nofootinbib]{revtex4-1}

\usepackage{mathtools}
\usepackage{xcolor}
\usepackage{amsfonts}
\usepackage{appendix}
\usepackage{amssymb}
\usepackage{graphicx}
\usepackage{srcltx}
\usepackage{catchfilebetweentags}
\usepackage{cancel}
\usepackage{siunitx}
\usepackage{fullpage}
\usepackage{booktabs}
\usepackage{todonotes}
\usepackage{footnote,ulem}
\usepackage{mathtools}
\usepackage{xcolor}
\usepackage{todonotes}
\usepackage{amssymb}
\usepackage{graphicx}
\usepackage{booktabs}
\usepackage{csquotes}

\usepackage{footnote}
\usepackage{braket}
\usepackage{siunitx}
\usepackage[english]{babel}
\usepackage{appendix}

\usepackage[T1]{fontenc}
\usepackage{lmodern}

\newcommand\numberthis{\addtocounter{equation}{1}\tag{\theequation}}

\DeclareTextFontCommand{\emph}{\it}

\begin{document}

\title{High-Visibility Time-Bin Entanglement for Testing Chained Bell
Inequalities}

%%%%%%%%%%%%%%%%%%%%%%%%%%%%%%%%%%%%%%%%%%%%%%%%%%%%%%%%%%%%%%%%%%%%%%%%%%%

\author{Marco Tomasin}
\affiliation{Department of Information Engineering, University of Padova, I-35131 Padova, Italy}
\affiliation{Istituto di Fotonica e Nanotecnologie, CNR, Padova, Italy}
\author{Elia Mantoan}
\affiliation{Department of Information Engineering, University of Padova, I-35131 Padova, Italy}
\affiliation{Istituto di Fotonica e Nanotecnologie, CNR, Padova, Italy}
\author{Jonathan Jogenfors}
\affiliation{Institutionen  f\"or systemteknik,  Link\"opings  Universitet,
SE-58183 Link\"oping, Sweden.}
\author{Giuseppe Vallone}
\affiliation{Department of Information Engineering, University of Padova, I-35131 Padova, Italy}
\affiliation{Istituto di Fotonica e Nanotecnologie, CNR, Padova, Italy}
\author{Jan-\AA{}ke Larsson}
\affiliation{Institutionen  f\"or systemteknik,  Link\"opings  Universitet,
SE-58183 Link\"oping, Sweden.}
\author{Paolo Villoresi}
\affiliation{Department of Information Engineering, University of Padova, I-35131 Padova, Italy}
\affiliation{Istituto di Fotonica e Nanotecnologie, CNR, Padova, Italy}

%%%%%%%%%%%%%%%%%%%%%%%%%%%%%%%%%%%%%%%%%%%%%%%%%%%%%%%%%%%%%%%%%%%%%%%%%%%

%%%%%%%%%%%%%%%%%%%%%%%%%%%%%%%%%%%%%%%%%%%%%%%%%%%%%%%%%%%%%%%%%%%%%%%%%%%

\begin{abstract}
    \noindent
    The violation of Bell's inequality requires a well-designed experiment to
    validate the result. In experiments using energy-time and time-bin
    entanglement, initially proposed by Franson in 1989, there is an intrinsic
    loophole due to the high postselection. To obtain a violation in this type
    of experiment, a chained Bell inequality must be used. However, the local
    realism bound requires a high visibility in excess of 94.63 percent in the
    time-bin entangled state. In this work, we show how such a high visibility
    can be reached in order to violate a chained Bell inequality with 6, 8 and
    10 terms.
\end{abstract}
%
%%%%%%%%%%%%%%%%%%%%%%%%%%%%%%%%%%%%%%%%%%%%%%%%%%%%%%%%%%%%%%%%%%%%%%%%%%%

%03.65.Ud: Entanglement and quantum nonlocality
%(e.g. EPR paradox, Bell's inequalities, GHZ states, etc.)
%03.67.Mn: Entanglement production, characterization, and manipulation
%42.50.Xa: Optical tests of quantum theory

\maketitle

%%%%%%%%%%%%%%%%%%%%%%%%%%%%%%%%%%%%%%%%%%%%%%%%%%%%%%%%%%%%%%%%%%%

\section{Introduction}
In his well known work of 1989,~\cite{fran89prl} Franson proposed a Bell
inequality for energy-time entanglement to investigate local realism. The
proposed experimental configuration consists of a source emitting two correlated
photons that are generated by a spontaneous parametric down conversion (SPDC)
process with a continuous-wave (CW) pump laser. The two photons are
directed towards two measurement stations, each consisting of identical
unbalanced Mach-Zehnder interferometers with path-length difference $\Delta L$.

The measurement stations can either impose or not impose a time delay $\Delta
L/c$ depending on the local phase setting.  Because of this, photon detection
may coincide in time, or be delayed on one side with respect to the other. The
optical path difference $\Delta L$ satisfies $\tau_c\ll\Delta L/c\ll\tau_p$
where $\tau_c$ and $\tau_p$ are the coherence time of the SPDC photons and the
pump, respectively. The lower bound ensures there is no first-order interference
of single photon counts and the upper bound ensures second-order interference
for coincident photon counts.

From all two-photon detections, only the 50\% coincident detections correspond
to entangled photons. Indeed, by postselecting those events where both photons
are detected within the time window $\Delta L/c$, two photon interference occurs
and a Bell inequality can be violated, modulo the postselection loophole
discussed in section\nobreakspace \ref {sec:theory}.

A different way of performing the experiment is by using time-bin entanglement.
This method differs from the standard Franson setup in that it uses a pulsed
laser pump and an additional unbalanced interferometer in the source. Here, the
photon pairs can be generated from the same pump pulse at two different moments
in time. In this case coherence is ensured by using a pump interferometer rather
than having to rely on the pump to be coherent, as would have been the case for
a CW pump without the interferometer.

As pointed out in several works~\cite{lars98pra,cabe09prl,joge14jpa,joge15sad},
Franson's scheme suffers from an intrinsic loophole because of the
aforementioned postselection. Which events that are discarded could in principle
be influenced by the measurement settings, and indeed, a local hidden variable
(LHV) model can reproduce the quantum correlations for the second order
interference exploiting the discarded
events~\cite{joge15sad,cabe09prl,lars98pra,deca94pra}. The main issue is that
the postselection is \enquote{nonlocal} in the sense that it requires
communication between the observers to know which events should be discarded.

Different approaches have been proposed to reach a violation of local realism
avoiding the postselection loophole. As  described
by~\cite{cabe09prl,joge14jpa,joge15sad} there are ways of re-establishing the
security of the Franson setup, including modifying the experimental
setup~\cite{cabe09prl,lima10pra,vall11pra,cuev13nco,carv15prl}. Additional
assumptions on the underlying physics have also been discussed by
Franson~\cite{fran99pra}; these, however, have given rise to new, undesirable
loopholes~\cite{lars98pra,joge14jpa}.

A consequence of using a time-bin entanglement setup instead of energy-time is
that it introduces a local postselection. The generated photon can be detected
only at three possible times ($t_0-\Delta L/c,t_0,t_0+\Delta L/c$) due to the
pulsed laser pump: the three arrival times will be denoted as while
\enquote{early} (E), \enquote{medium} (M) and \enquote{late} (L). Each observer
can perform a local postselection by discarding all events except those
occurring at $t_0$.

As will be shown later, such local postselection is not
sufficient for a loophole-free violation of the Bell test. Instead, only
\SI{50}{\percent} of all locally-postselected events correspond to a coincidence
and another, extra, postselection step is then required. This extra step
requires communication between Alice and Bob and is therefore nonlocal. As we
will show below, the LHV model~\cite{aert99prl,cabe09prl} previously introduced
for energy-time entanglement can be slightly modified to obtain an LHV model
also for time-bin entanglement.

In this work we use fast switching and so-called chained Bell inequalities in
order to obtain a violation of local realism in a time-bin experiment. In order
to facilitate this, we have designed and realized a source of time-bin entangled
photons with high visibility. It was previously believed that the visibility
requirements on such an experiment would be highly demanding, but we can
demonstrate visibilities up to \SI{99}{\percent} that make such a violation
possible.

\section{Theory}\label{sec:theory}
In a time-bin Bell experiment, a source device generates two time-correlated
photons, of which one is sent to Alice and one to Bob. Alice and Bob each randomly
select a measurement from the sets $\{A_i\}$ and $\{B_j\}$, respectively. Here,
$0\leq i,j \leq N$ for $N\geq 2$.

The pump interferometer is unbalanced with a difference in path length of
$\Delta L$. This leads to two distinct possibilities for when the photons are
emitted. We call the short-short (SS) events the photons generated by the pump
pulse traveling along the short arm, while long-long (LL) events are photons
generated by the pump pulse traveling along the long arm.

The measurement stations consist of unbalanced Mach-Zehnder interferometers with
two \enquote{paths} whose lengths differ by the same value $\Delta L$. Note that
\enquote{path} is written in quotation marks since one can't force particles to
have path realism without significantly altering the interpretation of the
outcomes~\cite{joge14jpa}. Instead, we view the effect of the measurement
station as having either a long or short delay on the particle. Then, due to the
pump interferometer and the measurement interferometer, each photon can be
detected at three possible times: ($t_0-\Delta L/c,t_0,t_0+\Delta L/c$).

If Alice and Bob perform measurements $A_i$ and $B_j$, the outcome ($-1$, $+1$)
and the time of detection is recorded at either end. After a number of trials,
Alice and Bob compare their measurement settings and time of arrival, and
whenever both photons are detected at $t_0$, that trial is used to compute the
quantum correlation $\langle A_i B_j\rangle$.
These correlations are then used to compute a statistical measure, for instance
the Clauser-Horne-Shimony-Holt (CHSH) value, which uses $N=2$ measurement
settings:
\begin{align}
    S_{\rm CHSH}=&\langle A_1 B_1\rangle-\langle A_2 B_2\rangle+\langle A_2
    B_1\rangle+ \langle A_1 B_2\rangle.
    \label{eqn:chsh}
\end{align}
In a classical system, $S_{\rm CHSH}$ has an upper bound of 2~\cite{clau69prl}.

As explained above, the \emph{time of arrival} is critical for each trial and is
therefore recorded by both Alice and Bob. According to Cabello et
al.~\cite{cabe09prl} and Aerts et al.~\cite{aert99prl}, a trial with differing
time of arrival between Alice and Bob has no quantum interference, which means
it must be discarded. Many events of this kind can be discarded locally, simply
by considering only the events in which the photon arrived at $t_0$. However, if
Alice and Bob were to only keep the photons arriving at $t_0$, then a detection
on Alice's side will have a \SI{50}{\percent} chance of not corresponding to a
detection on Bob side and vice versa.

In addition, a further \enquote{nonlocal} postselection is required to weed out
non-coincident events. This extra step is an inherent feature of time-bin
entanglement experiments, and on average, $50\%$ of the locally selected data
will be discarded here. As a consequence, the subensemble of finally selected
events becomes dependent on the phase settings, which must be considered when
choosing an appropriate Bell test for the experiment. With this dependence on
phase settings, it becomes possible to mimic the quantum-mechanical results
using a purely classical setup.

\begin{figure*}
    \centering
    \includegraphics[width=1\textwidth]{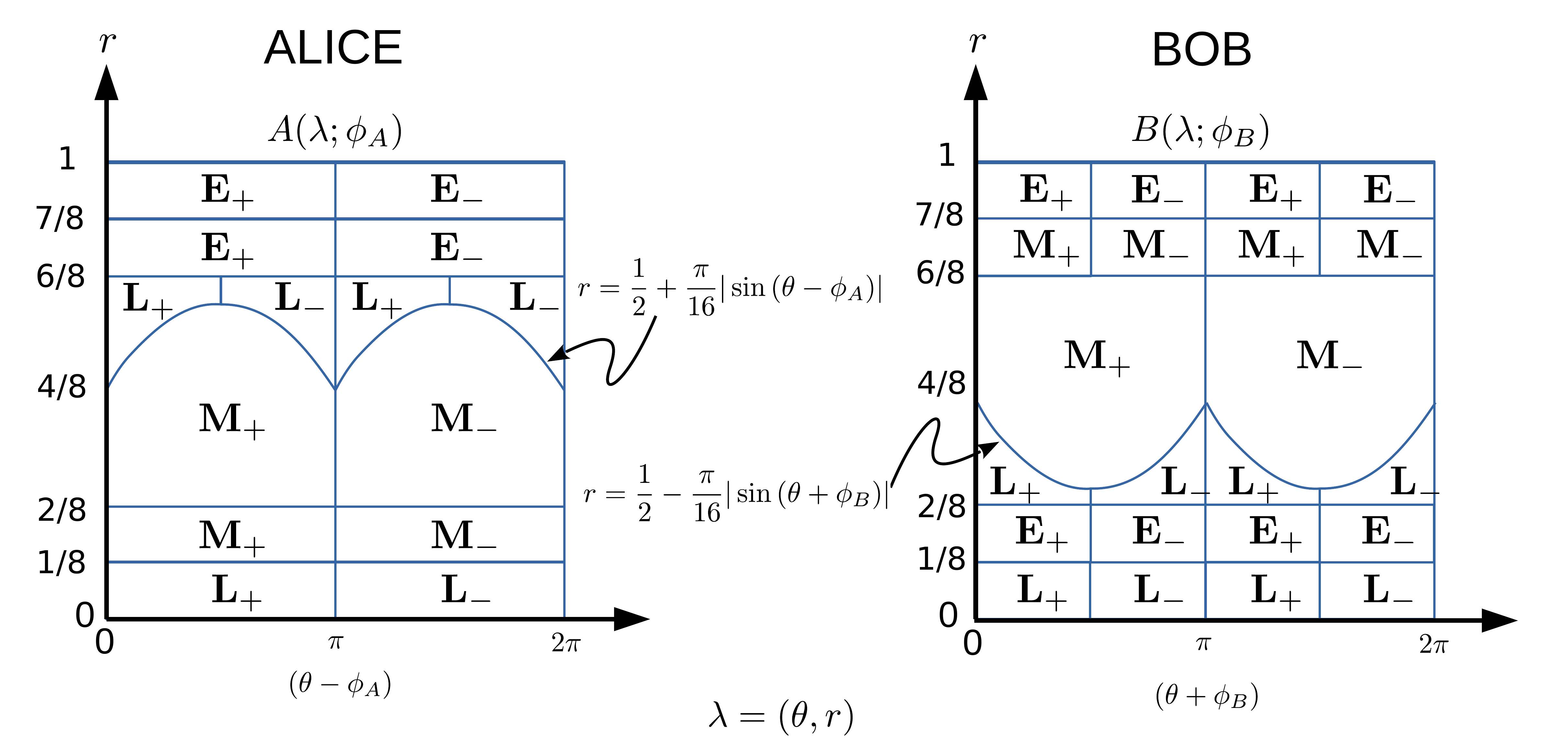}
    \caption{LHV model for time-bin entanglement.
        The hidden variable is a pair of numbers $\lambda=(\theta,r)$ uniformly
        distributed over the rectangle $\Lambda=\{(\theta,r):0\le\theta<2\pi,0\le
        r<1\}$. The outcomes, $\{{\rm E}_\pm, {\rm M}_\pm, {\rm L}_\pm\}$ are
        determined by the above graphs, with $\pm$ corresponds to the possible
        outcomes and while \enquote{early} (E), \enquote{medium} (M) and
        \enquote{late} (L) to the
        detection time. Indeed, E (M) events correspond to a photon generated from
        the first (second) pump pulse and travelling through the short (long) arm of
        the measuring interferometer. M events correspond to a photon generated from
        the first pump pulse and travelling through the long arm of the measuring
        interferometer or vice versa (second pump pulse and short arm).
        This model fully reproduces the quantum
    predictions for the time-bin setup.\label{fig:LHVtimebin}}
\end{figure*}

As shown in~\cite{joge14jpa}, if local realism holds in a time-bin entanglement
experiment, the outcomes obey
\begin{equation}
    S_{\rm re}\le3
    \label{eqn:time-bin-chsh-inequality}
\end{equation}
and we note that the value predicted by quantum mechanics, $S_{\rm qm}=2\sqrt2$
is \emph{not} in violation of inequality\nobreakspace \ref {eqn:time-bin-chsh-inequality}. In other
words, a local-hidden-variable (LHV) model such as fig.\nobreakspace \ref {fig:LHVtimebin} can
give the same outcome statistics as a truly quantum-mechanical experiment, as
predicted by~\cite{joge14jpa} and experimentally demonstrated
by~\cite{joge15sad}. This is called the \emph{postselection
loophole}~\cite{Larsson2014Loopholes}.

For applications such as Quantum Key Distribution (QKD) it is highly desirable
to re-establish a violation of a Bell inequality in the time-bin case. In order
to achieve this we must eliminate the postselection loophole which in turn
eliminates any LHV models such as the one in fig.\nobreakspace \ref {fig:LHVtimebin}.

A possible way forward is to employ \emph{fast switching} of the phase settings
and postselect the coincidence events in which both photons arrived at $t_0$. In
this case, the Bell inequality holds for the LL events, while a trivial bound is
obtained for the SS events (see~\cite{joge14jpa}). Even when taking this into
account, the standard Bell inequalities are insufficient as the quantum-mechanical
prediction still falls short of the corresponding inequality~\cite{joge14jpa}.

However, a violation can be obtained~\cite{joge14jpa} by using a generalized,
chained Bell-type inequality.  Here, we consider $N\geq 2$ measurements for each
observer so that the chained Bell parameter is given by
\begin{align}
    \notag
    S^{N}=&\langle A_{N}B_N\rangle+\sum^N_{k=2}[
    \langle A_k B_{k-1}\rangle+\langle A_{k-1} B_{k}\rangle]
    \\
    &-\langle A_1 B_{1}\rangle
    \label{eqn:generalized-inequality}
\end{align}
If $A_i$ and $B_j$ are phase measurement operators, the set of phases that
maximize eq.\nobreakspace \textup {(\ref {eqn:generalized-inequality})} are in consecutive
measurements separated by an angle of $\theta=\pi/(2N)$. The quantum prediction
in this case is
\begin{equation}
    S_{\rm QM}^{N}=2N \cos(\frac{\pi}{2N})
\end{equation}
with a classical bound of $2N-2$~\cite{brau90anp,joge14jpa}. In theory, such a
chained inequality can provide a strong violation but was
believed~\cite{joge14jpa,joge15sad} to be experimentally demanding to the point
of being unfeasible.

However, in a time-bin experiment, the realism bound is weaker than in a
standard Bell experiment (see Appendix\nobreakspace \ref {sec:chained} for details) and it is equal to
\begin{equation}
    S_{\rm LHV}^{N}=2N-1\label{eqn:chained-realism-bound}.
\end{equation}
The quantum prediction will be in violation of local realism if $S_{\rm
QM}^{N}>S_{\rm LHV}^{N}$ which can occur for $2N\ge6$. The bound in
eq.\nobreakspace \textup {(\ref {eqn:chained-realism-bound})} is due to the fact that the $SS$ events can be
influenced by the phase settings while the $LL$ subensemble is independent of
the phase setting~\cite{joge14jpa}.

Finally, there are purely experimental requirements on any Bell experiment. If
the experiment has lower than \SI{100}{\percent} visibility the bound in
eq.\nobreakspace \textup {(\ref {eqn:chained-realism-bound})} will be weakened so that local realism is not
in violation. For $2N\in \{6,8,10\}$ the critical visibility $V_{cr}= S_{\rm
LHV}/S_{\rm QM}^{N}$ is at least $94.63\%$~\cite{joge14jpa} which is
experimentally demanding.

As detailed in Appendix\nobreakspace \ref {sec:chained}, the inequality can be also written in the
CH-Hardy form involving only probabilities as
\begin{equation}\
    \begin{aligned}
        S_{\rm CH}^{N}=&p(a_{N}b_N)-
        \sum^N_{k=2}[p(a_{k}\bar b_{k-1})+
        p(\bar a_{k-1}b_k)]
        \\
        &-p(a_1b_1)\le 0
        \label{eqn:ch-hardy-form}
    \end{aligned}
\end{equation}
where $p(a_{i}b_j)$ is the joint probability of measuring $+1$ on Alice and Bob
side respectively and the bar indicates a $-1$ outcome. We note that
$S^N=4S_{\rm CH}^{N}+2(N-1)$. For time-bin entanglement, the correct LHV bound
is
\begin{equation}
    \frac34-N\leq S_{\rm CH}^{N}\leq\frac{1}{4}
\end{equation}
which is violated by the quantum prediction
\begin{equation}
    S^{N}_{{\rm CH},{\rm QM}}=\frac12-N\sin^2\frac{\pi}{4N}
\end{equation}
when $N\geq 3$.

\section{Experimental setup}
\begin{figure}
    \centering
    \includegraphics[width=7cm]{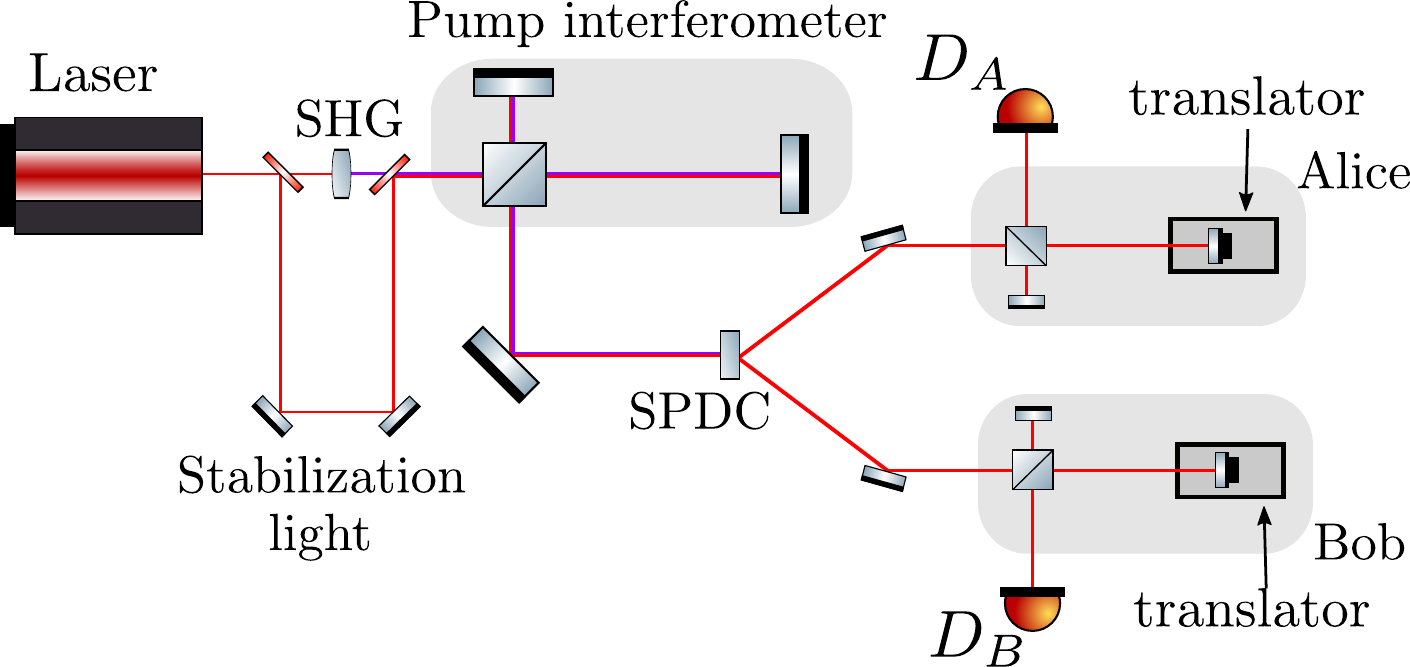}
    \caption{Setup of the time-bin entanglement experiment.\label{fig:setup}}
\end{figure}

As mentioned in section\nobreakspace \ref {sec:theory}, previous works~\cite{joge14jpa,joge15sad}
described the chained Bell inequality as experimentally demanding as it requires
a visibility in excess of \SI{94}{\percent}. In this section, however, we give
details of an experiment, depicted in fig.\nobreakspace \ref {fig:setup}, that gives us a high enough visibility to have a
working implementation of a chained Bell test in a time-bin entangled Franson experiment.

Obviously, this requires a very stable and precise measurement setup. The laser
source used in our experiment is a pulsed laser with a wavelength of
\SI{808}{\nano\meter}, \SI{76}{\mega\hertz} repetition rate and
$\approx$ \SI{150}{\femto\second} pulse width coming from a Ti:Sa mode-locked
oscillator, with an average power of \SI{2}{\watt}. The Ti:Sa oscillator beam is
used to pump a SHG crystal, which generates pulses at \SI{404}{\nano\meter}.

These pulses pass through a free-space
unbalanced Michelson interferometer (pump interferometer) where the light
\enquote{takes} the short, $\ket{S_p}$, or the long, $\ket{L_p}$ path and
finally pumps a SPDC crystal. As previously described in section\nobreakspace \ref {sec:theory}, the
length difference between the short and long paths is $\Delta L$.
The output of the source is photons with a wavelength of \SI{808}{\nano\meter}
that are injected into two measurement stations, one for Alice and one for Bob.

Each measurement station consists of a free-space unbalanced Mach-Zehnder
interferometer with a path length difference of $\Delta L$, matching the path
length difference of the source device. The mirrors on the long paths are
mounted on a piezoelectric translator in order to stabilize and change the phase
of the interferometers. Finally, the photons are coupled into single mode fiber
and directed toward two single photon avalanche photodiodes (SPADs). A
time-to-digital converter (TDC) with \SI{81}{\pico\second} resolution is used to
tag each detection event.

The generated state corresponds to
\begin{equation}
    \label{eqn:bell-state}
    |\Psi_e\rangle= \frac{1}{\sqrt{2}}\left(|S_{A}S_B\rangle+|L_{A}L_B\rangle \right).
    \numberthis
\end{equation}
Maintaining a constant phase difference between the two events
$|S_{A}S_B\rangle$ and $|L_{A}L_B\rangle$ is an important step in achieving high
visibility in eq.\nobreakspace \textup {(\ref {eqn:bell-state})}.

In order to achieve a fixed phase difference, Alice's and Bob's phases are
actively stabilized to the pump phase. This stabilization uses a small fraction
of the original oscillator beam and we call this the \emph{stabilization light}.
This stabilization light is injected into the pump interferometer after an
appropriate delay in order to prevent a detection overlap between the SPDC
photons. The outgoing light is split into two beams and each follows the same
path of the two SPDC photons. In this way, the second-order interference
generated through the pump and the measurement stations can be used to stabilize
and compensate the phase.

The interference pattern depends on the received photons. Therefore, the
measurement stations are simultaneously stabilized independently of each other.
Piezoelectric positioners with a resolution of \SI{1}{\nano\meter} control the
phase by measuring a feedback sensor. A measurement takes \SI{3}{\second}, after
which a stabilization process takes over, requiring roughly \SI{1}{second}
before starting over with the measurement procedure. The exact time required by
the stabilization depends on the number of single events.

Our stabilization allows us to stabilize not only the phase mismatch introduced
in the paths, but also the phase mismatch due to the pump laser. This is
important as the wavelength of the pump laser fluctuates due to variations in
the environmental room temperature which affects the phase and therefore the
visibility of the state.

\begin{figure}
    \centering
    \hspace*{-0.8cm}
    \includegraphics[width=0.57\textwidth]{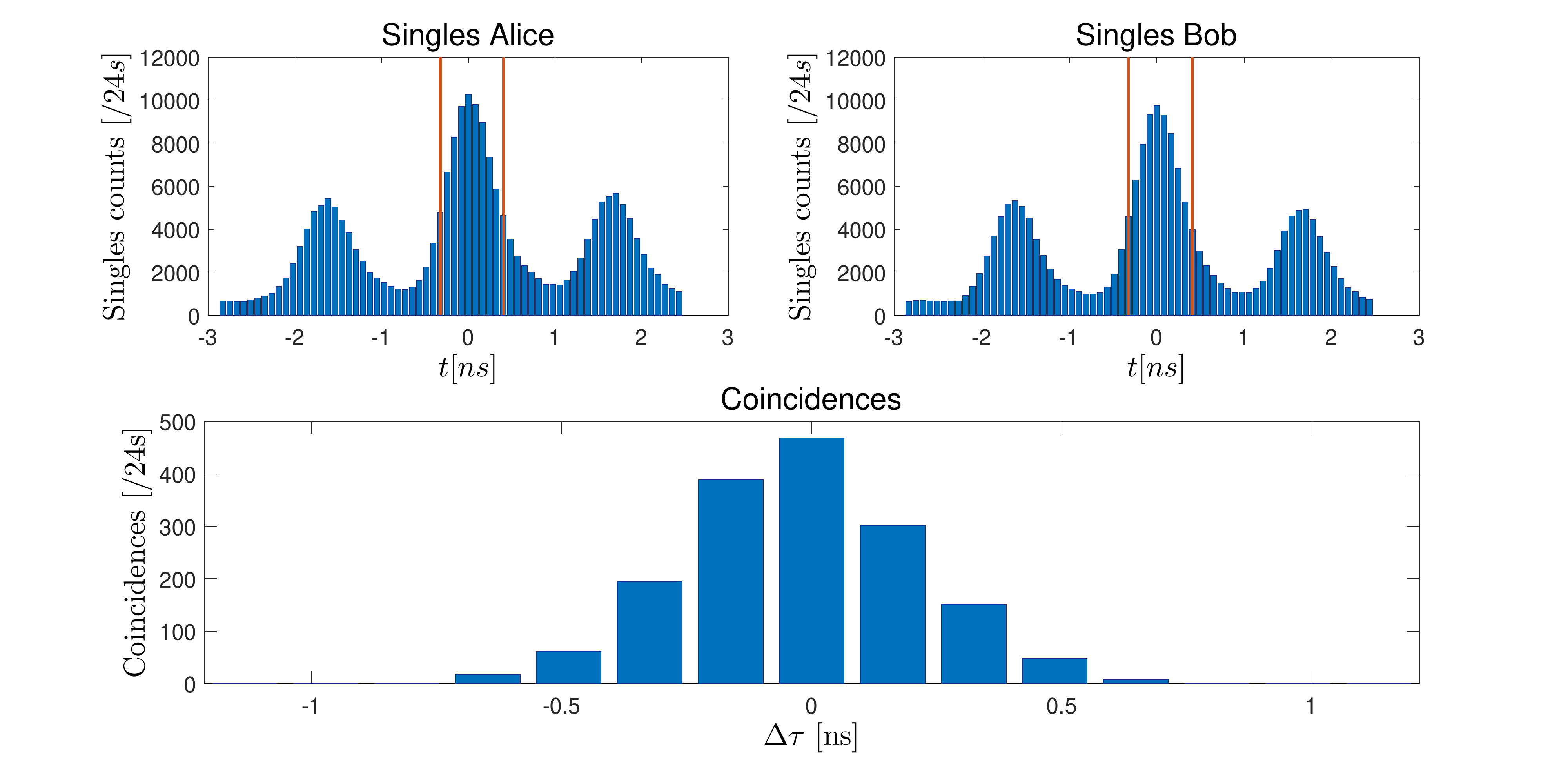}
    \caption{Time distribution of singles and coincidence count. The red bar on
        the singles graph represents the time interval within we consider an
        event to be coincident. The coincidences graph represents the
        coincidences between single events within the red bars, in a window of
    \SI{0.81}{\nano\second}. }\label{fig:all}
\end{figure}

\section{Experimental results}
Our experimental results are presented in fig.\nobreakspace \ref {fig:all} which shows the
distribution of single counts registered by Alice's and Bob's detectors together
with the joint coincidence. Thanks to the pulsed pump laser it is possible to
predict the arrival time of every generated photon at the detector. As expected,
the single counts are concentrated around three distinct peaks, where the
central peak corresponds to entangled events.

The coincidence window is indicated by the red bars in fig.\nobreakspace \ref {fig:all} and is
determined \emph{a priori} by using the emission time of the pump laser beam as
a reference. The horizontal axis in the singles plots of fig.\nobreakspace \ref {fig:all}
represents the arrival time of the Alice (or Bob) photon with respect to this
reference. Here, $t=0$ is the detection corresponding to $\ket{L_{A}L_B}$ and
$\ket{S_{A}S_B}$ events. The coincidence window therefore matches $|t|\leq
\SI{0.405}{\nano\second}$, corresponding to $\pm 5$ bins of the TDC converter.

An event at Alice's measurement station is said to be coincident with an
event at Bob's measurement station if they fall inside this coincidence window.
Figure\nobreakspace \ref {fig:all} shows the distribution of coincidences as a function of the time difference
$\Delta\tau$ between Alice's and Bob's events. Note that, by definition,
$|\Delta\tau|$ is smaller that the detection window $|\Delta
\tau|\leq \SI{0.810}{\nano\second}$.
\begin{figure}
    \centering
    \hspace*{-0.8cm}
    \includegraphics[width=0.57\textwidth]{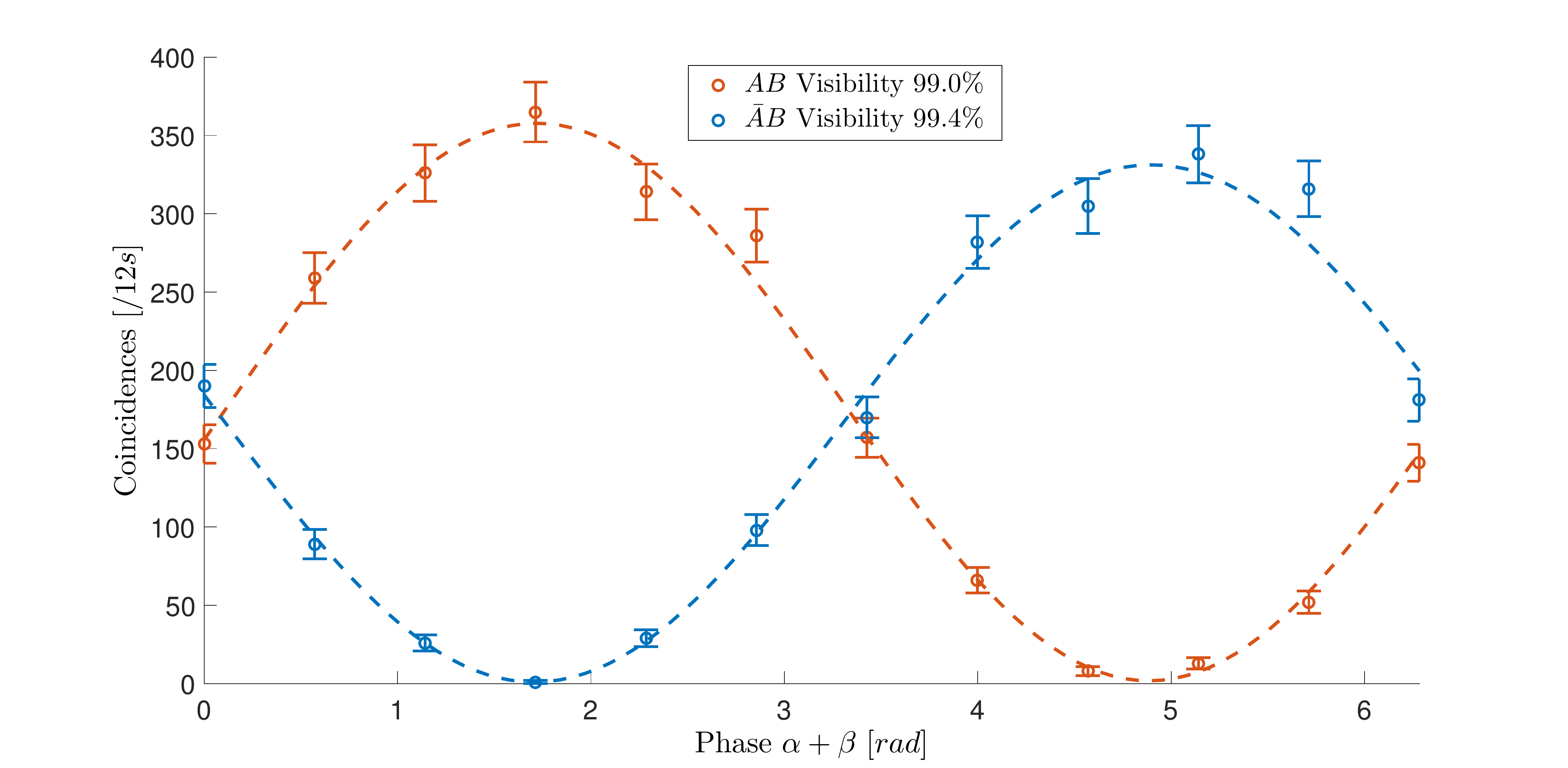}
    \caption{Raw coincidence counts between Alice's and Bob's detectors for different
        values of phases (red data). The blue data represents the coincidence
counts obtained by shifting
Alice's phase by $\pi$ with respect to the red data.\label{fig:doub}}
\end{figure}

Our experiment uses only one detector each for Alice and Bob. Therefore,
four measurements are required to calculate the correlation. This corresponds
to a measurement in the standard Franson interferometer if we assume that the correlation
obtained with such a measurement is equal to the correlation obtained with two
detectors in a single measurement.

The experiment was performed three times, corresponding to a chained Bell
inequality using $2N=6$, $8$ and $10$, respectively. In the case of $2N=6$,
choosing $\theta=\pi/6$ yields the quantum-mechanical prediction $S_{\rm
CHSH}^3=5.163$. For $2N=8$, $\theta=\pi/8$ yields the corresponding prediction
$S_{\rm CHSH}^4=7.169$ and finally $2N=10$ with choosing $\theta=\pi/10$ yields
$S_{\rm CHSH}^{5}=9.271$ in the same way. The measured results from these three
experiments are shown in table\nobreakspace \ref {tab:results}.
\begin{table}
    \centering
    \begin{tabular}{cp{0.5cm}cp{0.5cm}cp{0.5cm}cp{0.5cm}cl}
        $i$&&$S^{3}_{\rm LHV,i}$&&$S^3_{\rm CH,i}$&&$err_S$&&Violation \\
        \hline
        \hline
        1&&0.25&&0.289&&0.011&&3.40$\sigma$\\
        2&&-2.25&&-2.335&&0.020&&4.15$\sigma$\\
        3&&-2.25&&-2.247&&0.020&&-0.17$\sigma$\\
        4&&0.25&&0.293&&0.012&&3.66$\sigma$\\
        \bottomrule[1.5pt]
        \rule{0pt}{10pt}
        $S^3_{\rm CHSH}$&&5&&5.163&&0.033&&4.91$\sigma$\\
    \end{tabular}

    \vspace*{20pt}

    \begin{tabular}{cp{0.5cm}cp{0.5cm}cp{0.5cm}cp{0.5cm}cl}
        $i$&&$S^{4}_{\rm LHV,i}$&&$S^4_{\rm CH,i}$&&$err_S$&&Violation \\
        \hline
        \hline
        1&&0.25&&0.282&&0.011&&3.06$\sigma$\\
        2&&-3.25&&-3.299&&0.022&&2.28$\sigma$\\
        3&&-3.25&&-3.284&&0.022&&1.59$\sigma$\\
        4&&0.25&&0.302&&0.011&&4.94$\sigma$\\
        \bottomrule[1.5pt]
        \rule{0pt}{10pt}
        $S^4_{\rm CHSH}$ &&7&&7.169&&0.034&&4.93$\sigma$\\
    \end{tabular}

    \vspace*{20pt}

    \begin{tabular}{cp{0.5cm}cp{0.5cm}cp{0.5cm}cp{0.5cm}cl}
        $i$&&$S^{5}_{\rm LHV,i}$&&$S^{5}_{\rm CH,i}$&&$err_S$&&Violation \\
        \hline
        \hline
        1&&0.25&&0.307&&0.009&&6.76$\sigma$\\
        2&&-4.25&&-4.304&&0.022&&2.64$\sigma$\\
        3&&-4.25&&-4.331&&0.021&&4.01$\sigma$\\
        4&&0.25&&0.327&&0.009&&8.89$\sigma$\\
        \bottomrule[1.5pt]
        \rule{0pt}{10pt}
        $S^{5}_{\rm CHSH}$&&9&&9.271&&0.031&&8.67$\sigma$\\
    \end{tabular}
    \caption{Results for three different sets, with $N= 3,\,4,\,5$. $S_{\rm
        LHV,i}^N$ correspond to the value predicted by a LHV model. The reported
        $S^N_{\rm CH,i}$ has been calculated from the raw coincidence by using
        eq.\nobreakspace \textup {(\ref {eqn:sch1})} and eq.\nobreakspace \textup {(\ref {eqn:sch2})}, while the $S^N_{\rm CHSH}$ are obtained
    from eq.\nobreakspace \textup {(\ref {eqn:Schsh4})}.\label{tab:results}}
\end{table}

As for visibility, our stabilization setup has good performance as seen in
fig.\nobreakspace \ref {fig:doub}. By measuring the coincidence as a function of the phase
$\alpha+\beta$ we obtain a measured visibility up to $99\%$ which is more than
suitable for the chained Bell inequality.

\section{Conclusions}
Device independence is a powerful theoretical framework where many of the usual
complications in applications such as QKD are reduced to a single statistical
Bell test. Instead of a painstakingly complex interpretation where every
component, path, defect etc.\ impacts the physical interpretation of the
experiment (for instance, the idea of \enquote{path realism} as discussed
by~\cite{joge14jpa}), device independence certifies the entire setup as
correct in a single step.

In addition, time-bin entanglement has a higher inherent noise
rejection~\cite{joge15sad} than traditional methods such as polarization
encoding. This can lead to simpler devices with less moving parts than required
by traditional protocols, opening the door for a wider variety of applications.

In this work we show that is possible to reach a high visibility in a
device-independent experiment using time-bin entanglement in order to violate a
chained Bell inequality. By using $2N\geq 6$ settings at each measurement
station, no local hidden variable model can reproduce the predictions of a
quantum experiments. The postselection loophole, which is present in the case
$2N=4$ (i.e.\ the standard CHSH-Bell inequality), is therefore avoided with the
generalized, chained, Bell inequality with $2N\geq 6$.

Note that in order to fully violate the chained Bell inequality, fast switching
must be used so that the phase settings at the measurement stations are randomly
chosen at least every $\Delta L/c$. This has not been performed by the current
experiment, but could be done by using a
fast phase modulator, synchronized with the pulse laser, to change the phase
within the interferometers. If $2N\geq 3$ is combined with fast switching, all requirements
set out by~\cite{aert99prl} are fulfilled.

Generally, chained Bell inequalities demand a high experimental visibility
($\gtrsim$ \SI{94.63}{\percent}). Our experiment fulfills this condition with a
visibility of \SI{99}{\percent} and shows that, despite the difficulties, it is
therefore experimentally possible to implement chained Bell inequalities. This
opens the door for applications such as device-independent quantum key
distribution based on time-bin entanglement without the weaknesses inherent in
the original Franson system.

\acknowledgments{}

\appendix
\section{Chained Bell inequalities with time-bin entanglement}\label{sec:chained}
In this section we show how to derive the chained Bell-inequality for time-bin
entanglement in two different forms, CH and CHSH. The chained Bell
inequalities are a generalization of the CHSH and Clauser-Horne (CH)
inequalities introduced by~\cite{pear70prd} and~\cite{brau90anp}:
such inequalities consider the scenario in which Alice and Bob choose among $2N
\geq 4$ dichotomic observables (with outputs $+1$ or $-1$).

The CH form of the chained Bell inequality is
\begin{equation}
    \label{CH_appx}
    1-N\leq S^N_{\rm CH}\leq 0
\end{equation}
where the Bell parameter is given by
\begin{equation}
    \begin{aligned}
        S^N_{\rm CH}=&p(a_{N}b_N)+\sum^N_{k=2}[p(a_{k}b_{k-1})+p(a_{k-1}b_k)]
        \\
        &-p(a_1b_1)-\sum^N_{k=2}[p(a_k)+p(b_k)]
    \end{aligned}
\end{equation}
In the above expression, $p(a_{i}b_j)$ is the joint probability of measuring $+1$
on Alice's and Bob's side, respectively. If we assume fair sampling, the single side
probabilities $p(a_k)$ and $p(b_k)$ can be expressed as
$p(a_k)=p(a_kb_{k-1})+p(a_k\bar b_{k-1})$ and $p(b_k)=p(a_{k-1}b_{k})+p(\bar
a_{k-1} b_k)$ where $\bar a_j$ means the $-1$ outcome. The above replacement
leads to
\begin{equation}\label{eqn:sch1}
    \begin{split}
        S^N_{\rm CH,1}=&p(a_{N}b_N)-\sum^N_{k=2}
        [p(a_k\bar b_{k-1})
        +p(\bar a_{k-1}b_k)]\\
        &-p(a_1b_1)
    \end{split}
\end{equation}
Since we can arbitrarily define the measurement outcomes with the $+1$ and $-1$
inverted, the inequality also holds when $a_k$ or $b_k$ are replaced by $\bar
a_k$ and $\bar b_k$, obtaining three other Bell parameters given by
\begin{equation}\label{eqn:sch2}
    \begin{aligned}
        S^N_{\rm CH,2}=&p(\bar a_N\bar b_N)
        -\sum^N_{k=2}[p(\bar a_{k}b_{k-1})
        +p(a_{k-1}\bar b_k)]\\
        &-p(\bar a_1 \bar b_1)
    \end{aligned}
\end{equation}
\begin{equation}\label{eqn:sch3}
    \begin{aligned}
        S^N_{\rm CH,3}=&p(a_N\bar b_N)-
        \sum^N_{k=2}[p(a_{k}b_{k-1})
        +p(\bar a_{k-1} \bar b_k)]\\
        &-p(a_1 \bar b_1)
    \end{aligned}
\end{equation}
\begin{equation}\label{eqn:sch4}
    \begin{aligned}
        S^N_{\rm CH,4}=&p(\bar a_{N}b_N)
        -\sum^N_{k=2}[p(\bar a_{k}b_{k-1})+p(a_{k-1}b_k)]\\
        &-p(\bar a_1 b_1)
    \end{aligned}
\end{equation}
By combining eqs.\nobreakspace  \textup {(\ref {eqn:sch1})} to\nobreakspace  \textup {(\ref {eqn:sch4})}  it is
possible to derive the inequality in the CHSH form given by
\begin{equation}
    \label{eqn:CHSH_appx}
    |S^N_{\rm CHSH}|\leq 2N-2,
\end{equation}
where
\begin{equation}
    \label{eqn:Schsh4}
    \begin{aligned}
        S^N_{\rm CHSH}=&S^N_{\rm CH,1}+S^N_{\rm CH,2}
        -S^N_{\rm CH,3}-S^N_{\rm CH,4}        \\
        =&\langle A_{N}B_N\rangle-
        \sum^N_{k=2}
        [\langle A_{k}B_{k-1}\rangle+
        \langle A_{k-1}B_k\rangle]  \\
        &-\langle A_1 B_1\rangle.
    \end{aligned}
\end{equation}
In eq.\nobreakspace \textup {(\ref {eqn:Schsh4})} the correlations are defined as
\begin{equation}
    \begin{split}
        \langle A_{k}B_j\rangle&=
        p(a_{k}b_j)+p(\bar a_k\bar  b_j)-
        p(\bar a_k b_j)-p(a_k \bar b_j)
        \\
        &=4p(a_{k}b_j)-2p(a_k)-2p(b_j)+1
        \label{eqn:correlations}
    \end{split}
\end{equation}

As discussed in section\nobreakspace \ref {sec:theory}, eq.\nobreakspace \textup {(\ref {eqn:CHSH_appx})} does not apply to
time-bin entanglement due to
the postselection of events. Instead, we must
distinguish between the early-early ($EE$) and the late-late ($LL$) events:
eq.\nobreakspace \textup {(\ref {CH_appx})} holds only for the LL events, while the
early-early (EE) events are only governed by the trivial inequality
\begin{equation}
    \label{CHSH_EE}
    |S^N_{\rm CHSH,EE}|\leq 2N
\end{equation}
corresponding to the following bound in the CH-form:
\begin{equation}
    \label{CH_EE}
    \frac12-N\leq S^N_{\rm CH,EE}\leq \frac12.
\end{equation}
The bound is trivial since $S^N_{\rm CHSH,EE}$ contains $2N$ terms and each
term has an absolute value $\leq 1$. Since $EE$ and $LL$ events equally
contribute to the coincidences, the right-hand side of the inequality is
therefore given by the average
of the $EE$ and $LL$ right-hand sides, namely $\widetilde S^N=\frac12S^N_{\rm
EE}+\frac12S^N_{\rm LL}$. We then get the following correct inequalities for time-bin
entanglement:
\begin{equation}
    \label{CHSH_EE2}
    |\tilde S^N_{\rm CHSH,EE}|\leq 2N-1
\end{equation}
and
\begin{equation}
    \label{CH_appx2}
    \frac34-N\leq \tilde S^N_{\rm CH}\leq \frac14
\end{equation}
We note that since $S_{\rm CH}$ involves only joint probabilities and not
correlation, the inequality is valid when we use one detector at each side
the measure the probabilities $p(a_{i}b_j)$. Since we have demonstrated that
the CHSH form can be derived from the CH form, it implies that also the CHSH
inequality holds when one detector at each side is used.

\section{LHV model for time-bin entanglement}
In this section we give the detail of the LHV model for time-bin entanglement.
Consider the general time-bin scheme, where each observer detect the photon at
three possible different arrival times: $t_0-\Delta T$  (here we call this event
early, E), $t_0$ (medium, M) and $t_0+\Delta T$  (late, L), where $\Delta
L=c\Delta T$ is the length difference between the short and long paths. For each
detection time, $E$, $M$ and $L$, a measurement indicated by $a$ and $b$ for
Alice and Bob, can have two outcomes, $+$ and $-$.

\noindent
Now consider the probability
\begin{equation}
    P(a,b|\phi_A,\phi_B)
\end{equation}
where $\phi_A$ and $\phi_B$ are the measurement settings for Alice and Bob.
For the detection $M-M$, quantum mechanics predicts the following probabilities:
\begin{align*}
    P(M_+,M_+|\phi_A,\phi_B)=&P(M_-,M_-|\phi_A,\phi_B)
    \\
    =&\frac{1}{16}\left[1+\cos(\phi_A+\phi_B)\right]
    \\
    P(M_+,M_-|\phi_A,\phi_B)=&P(M_-,M_+|\phi_A,\phi_B)
    \\
    =&\frac{1}{16}\left[1-\cos(\phi_A+\phi_B)\right].
\end{align*}
In the other cases, we have
\begin{align*}
    &P(M_\pm,E_\pm|\phi_A,\phi_B)=P(E_\pm,M_\pm|\phi_A,\phi_B)=\frac{1}{32}
    \\
    &P(M_\pm,L_\pm|\phi_A,\phi_B)=P(L_\pm,M_\pm|\phi_A,\phi_B)=\frac{1}{32}
    \\
    &P(E_\pm,E_\pm|\phi_A,\phi_B)=P(L_\pm,L_\pm|\phi_A,\phi_B)=\frac{1}{32}
    \\
    &P(E_\pm,L_\pm|\phi_A,\phi_B)=P(L_\pm,E_\pm|\phi_A,\phi_B)=0.
\end{align*}
A local hidden variable (LHV) model can mimic the above correlations in the
following way. Consider a hidden variable $\lambda=(\theta,r)$, with $\theta$ uniformly
distributed between $0$ and $2\pi$, and $r$ uniformly distributed between $0$
and $1$. The outcome $A(\lambda;\phi_A)$ and $B(\lambda;\phi_B)$ can then be
determined from fig.\nobreakspace \ref {fig:LHVtimebin} in the following way: Given a fixed value
$(\theta,r)$, the LHV model deterministically establishes the outcomes of Alice
and Bob as functions of $\phi_A$ and $\phi_B$ given by $A(\lambda;\phi_A)$ and
$B(\lambda;\phi_B)$.

This LHV model is local, because the outcomes of Alice do not
depend on $\phi_B$ and vice versa.
The probabilities predicted by the LHV model are:
\begin{align*}\label{eq:intLHV}
    \notag
    &P{(a,b|\phi_A,\phi_B)}_{LHV}
    &=\frac{1}{2\pi}\int_{S(a,b,\phi_A,\phi_B)}\textrm{d}\theta \textrm{d}r\,,
\end{align*}
where $S(a,b,\phi_A,\phi_B)$ is the subset of the possible hidden variable
values such that $A(\theta,r;\phi_A)=a$ and $B(\theta,r;\phi_B)=b$. In the above
equation the possible values of $a$ and $b$ are $\{E_\pm,M_\pm,L_\pm\}$ for
Alice and Bob. It is easy to check that the above probabilities coincide with
the predictions of quantum mechanics.

\bibliographystyle{prl_with_titles}
\bibliography{timebin}

\end{document}